\documentclass{article}
\usepackage{amsmath,amssymb}
\usepackage[preprint]{nips_2018}
\usepackage{graphicx}
\usepackage{svg}
\usepackage{booktabs} 
\usepackage{xcolor}
\usepackage{wrapfig}

\RequirePackage{snapshot}

\DeclareMathOperator*{\argmax}{argmax}
\DeclareMathOperator{\E}{\mathbb{E}}

\newcommand{\nmk}{{n {\scriptscriptstyle -} k}}

\usepackage{nicefrac}
\newcommand\bsfrac[2]{%
\scalebox{-1}[1]{\nicefrac{\scalebox{-1}[1]{$#1$}}{\scalebox{-1}[1]{$#2$}}}%
}

\usepackage{natbib}

\title{Learning multiagent coordination in the absence of communication channels}

\author{Aaron Goodman\\
  Department of Biology\\
  Stanford University\\
  Stanford, California\\
  \texttt{aaronjg@stanford.edu}
}

\begin{document}
  \maketitle
  
  \begin{abstract}
    In this work, we develop a reinforcement learning protocol for a multiagent coordination task in a discrete state and action space: an iterated prisoner's dilemma game extended into a team based, winner-take all tournament, which forces the agents to collude in order to maximize their reward. By disallowing extra communication channels, the agents are forced to embed their coordination strategy into their actions in the prisoner's dilemma game. 

    We develop a representation of the iterated prisoner's dilemma that makes it amenable to Q-learning. We find that the reinforcement learning strategy is able to consistently train agents that can win the winner take all iterated prisoner's dilemma tournament. 

    By using a game with discrete state and action space, we are able to better analyze and understand both the dynamics and the  communication protocols that are established between the agents. We find that the agents adapt a number of interesting behaviors, such as the formation of benevolent dictators, that minimize inequality of scores. We also find that the agents settle on a remarkably consistent symbology in their actions, such that agents from independent trials are able to collude with each other without further training.

  \end{abstract}
    
  \section{Motivation}
  
  In domains such as biology, economics, and politics, winner-take-all dynamics emerge when a single entity can secure enough resources in the short run to ultimately seize control of all available resources.  In simple microbial communities, slight growth advantages accumulate over time to allow the fittest to take over. In economics, economies of scale lead to cumulative advantage and winner-take-all markets~\cite{frank2013}. In politics, first-past-the-post voting systems concentrate political power in a single winner, regardless of their margin of victory. Unsurprisingly, these dynamics can lead to intense competition as entities strive to win both by improving themselves and by sabotaging their competitors.

  At the same time, competition is often mitigated by the capacity of any one individual to gain control of enough resources to completely out-compete their opponents. Thus, individuals have an incentive to band together to ensure that one of their own is the ultimate victor, with the spoils distributed among the team members. However, in order for such unions to form, the agents must coordinate with each other to establish a strategy.

  Coordinating the strategy can be difficult since interacting agents interacting do not have full visibility into each other's internal state. Thus, agents look for visible signals that are understood to correlate with coordinating behavior~\cite{hamilton1964}. Furthermore, in some environments, exogenous factors, such as regulators prohibit explicitly coordination, so agents wishing to collude must do so implicitly~\cite{Ohlhausen2017}.

  In this work, we will focus on how agents can embed a primitive ``language'' to establish coordination into their action space. We use language here in the sense of Lewis, as signals that have become coupled to the signified solely based on agreed upon convention~\cite{lewis1969}. Furthermore, we show that the signifiers in the language that emerges becomes coupled with the signified based on the environment, and obeys the properties of a rational exchange~\cite{horn1984}, and thus agents trained independently can still establish coordination.

To study dynamics in such complex systems, researchers rely on simplified models of dynamics. The iterated prisoner's dilemma (IPD) is a well-studied model for understanding the emergence of cooperation in biology and economics~\cite{Nowak1993} and is a natural starting point for our study of learning implicit communication.  Briefly, the standard prisoner's dilemma game presents two players with a choice: to cooperate or defect with their opponent. The total mutual benefit is maximized when both cooperate even though each individual has an incentive to defect. 

The iterated prisoner's dilemma is an extension of the one-shot prisoner's dilemma, where players play a series of bouts against one another. Unlike the one-shot case, the optimal strategy here is not so clear. Good strategies often allow for mutual cooperation and punishment of defection~\cite{Axelrod1981}.  However, the optimal strategy depends on the nature of the tournament. One particularly successful class of strategies drives players to identify other players following an identical strategy of themselves and then collude with each other. These strategies are typically carefully designed to have a specific handshake ~\cite{Prase2011}.

Rather than designing such communication strategies into our agents, we are interested in using reinforcement learning to discover such strategies, analyze the learning process to understand how such strategies could evolve, and explore the dynamics by modulating the structure of the game and the opponents to understand how different environments and reward structures permit different stable strategies. To explore this, we will use a winner-take-all round-robin, iterated prisoner's dilemma tournament.  We impose a winner-take-all tournament structure, in which the winner of the tournament receives the cumulative payoff from all of the agents in order to entice collusion between agents. We further employ a round-robin tournament structure so that each agent faces every other agent exactly ounce in the tournament.

Thus we are interested in understanding how such collusion relationships could develop via \textit{implicit} communication. That is, communication that occurs through the actions that are directly tied to the payoffs themselves, rather than over a separate communication channel.  The iterated prisoner's dilemma is a useful model for studying such questions due to the emergent dynamics that can arise from simple rules.

\subsection{Related Work}
Previous work in developing learning algorithms for the iterated prisoner's dilemma included using neural networks to learn about the opponent within tournaments~\cite{Seiffertt2009}, learning strategies for evolutionary algorithms~\cite{Harrald1996}, or training generic agents that can learn many games~\cite{Zawadzki2008}.  Our work differs from these in two key regards. First, we are attempting to establish a tournament policy, rather than a per game policy. Thus what other papers refer to as within tournament adaptive learning, we would formulate as a single policy in our analysis. Second, previous work develops strategies that seek to achieve the highest payout for an individual agent in a tournament~\cite{Stewart2012,harper2017}, whereas our focus on winner takes all tournaments finds an optimal strategy that maximizes the payouts of all agents in the tournament, conditioned on the focal agent winning the tournament.

Other work has explored multi-agent cooperation and the evolution of communication protocols that allow agents to share information, in terms of centralized and localized learning using Deep Q Learning. In the centralized approach, agents are able to leverage communication protocols to backpropagate error derivatives~\cite{foerster2016,Sukhbaatar2016}. Other approaches that have investigated the case of narrow communication channels assumed cooperation and used multiagent coordination to learn policies that allow the agents to work together to maximize the shared objective function~\cite{Melo2012,Zhang2016}.

Work using continuous time and space social games found that cooperation emerges in the presence of abundant resources, but reduces if there is a shortage of global resources, leading to conflict~\cite{Leibo2017,Lowe2017}, and that changing incentive structures in the game can promote or inhibit cooperation~\cite{Tampuu2015}.The nature of the continuous time and space facilitates the development of gradient-based learning algorithms. However, the continuous state spaces make it difficult to understand the actual communication strategies learned by the agents.

\section{Winner-Take-All Iterated Prisoner's Dilemma}

\begin{wraptable}{r}{5.5cm}
  \centering
  \caption{Payoff matrix for a single round of Iterated Prisoner's Dilemma}
  \label{payoff}

\begin{tabular}{l c c}
\toprule
\bsfrac A B&  Collude & Defect \\
\midrule
  Collude & \bsfrac 3 3 & \bsfrac 5 0 \\
  Defect & \bsfrac 0 5 & \bsfrac 1 1 \\
\bottomrule
\end{tabular}

\end{wraptable}

In the one-shot prisoner's dilemma, the  payoff matrix is such that each agent has an incentive to defect, yet mutual cooperation leads to the highest payoff matrix. We use the same payoff matrix as in Axelrod's original tournament, which is given in Table \ref{payoff}~\cite{Axelrod1981}.

In the iterated prisoner's dilemma, two agents compete over several of bouts, with the final payoff being the sum of the payoffs over the rounds.  In the individual-based tournament, each agent plays a round-robin tournament against every other agent  exactly once, and the order of the matches is randomized prior to the start of the tournament. Players are able to see their opponent's past actions, but cannot observe any matches in the tournament that they do not participate in. They cannot observe the identity of their opponents, and no side channels of communication are allowed.

The tournament is set up in a winner-take-all fashion, such that at the end of the tournament, the agent with the highest score receives a reward as the cumulative payout from all players in all rounds of the tournament, while the losers receive a reward of 0.
However, tournaments may also be set up to with teams that divide the reward evenly in side payments after the tournament, thus encouraging collusion.  

\section{MDP formulation of Prisoner's Dilemma}
In this section we develop the round-robin, iterated prisoner's dilemma as a Markov decision process. Previous work on developing machine learning algorithms to find policies for iterated prisoner's dilemma has used feature engineering to condense the state space available to the agents~\cite{harper2017,franken2005}. This work has empirically yielded quite effective policies, but such restrictions on the state space visible to the agents unacceptably restricts the available communication policy.

\subsection{One Off Prisoner's Dilemma}
In the case of the one off prisoner's dilemma, a policy is either ``Cooperate'' or ``Defect''.
No matter what the opponents policy, ``Defect'' has a higher payoff.

In the one off prisoner's dilemma, there is a single starting state, $s_0$, and the action space $\mathcal{A} := \{C,D\}$.

\subsection{Iterated Prisoner's Dilemma}
In the iterated prisoner's dilemma, the action space remains the same, however the state space increases. We can define the state as the actions taken in the bouts up to the current round.   Each bout, $\mathcal{B}$ of the prisoner's dilemma has one of four outcomes, and the state space, $\mathcal{S}$ is made up of all the possible bouts, up to the tournament length, $n$.
\[
\mathcal{B} \in \left\{(C,C),\,(C,D),(D,C),\,(D,D) \right\}, ~~~~ ~~\mathcal{S} = \cup_{0\leq i \leq n} \mathcal{B}^i 
\]
Thus a policy function $\pi$ is a surjective map from states to actions: $\pi : \mathcal{S} \rightarrow \mathcal{A}$, and $\Pi$ is the set of all such maps. We define $r_s$, as the cumulative reward earned by the agent in state $s$ as the sum of the reward for each bout in $s \in \mathcal S$.  Furthermore, for convenience we define a restricted state space of the states possible after bout $k$ of the tournament to be $\mathcal S_k = \mathcal B^k$

Then also define $P(s_k| \pi_1,\pi_2)$ as the probability of arriving at $s_k$ given the opponent is using policy $\pi_2$ and we are using policy $\pi_1$, and we can then find $P(\pi_2|s_k,\pi_1)$ using Bayes rule.

We define the value of the policy, $\pi$, from the initial state $s_0$, against an agent playing policy $\pi'$ to be 
\begin{equation}
  V^{\pi}(s_0,\pi') = \E_{s\in \mathcal S_n}[r_s]  = \sum_{s_n \in \mathcal S_n} r_{s_n} P(s_n | \pi,\pi')
\end{equation}
Thus we can define the optimal policy, $\pi^*_i$, when faced against opponent $\pi_i$, to be $\pi^*_i = \argmax_{\pi \in \Pi} V^{\pi}(s_0,\pi')$.
  \subsection{IPD: Unknown Opponent}
  \label{sec:unknownopponent}
  In the iterated prisoner's dilemma tournament, the agent must find an optimal policy even though it  does not know the policy of it's opponent.
  
We could think about a na\"{i}ve strategy to do this would be to look at all of the optimal strategies against our prospective opponents, and pick the one that is the best overall, given the probability of encountering the opponent, $P(\pi')$. That is, $\Pi' = \{\pi \in \Pi ~|~ \pi = \pi_i^*  ~\text{for some}~ i \}$ and

\begin{equation*}
\pi^* = \argmax_{\pi \in \Pi'} \E[V^\pi(s_k,\pi')] =   \argmax_{\pi \in \Pi'} \sum_{\pi'} V^\pi(s_0,\pi') P(\pi')
\end{equation*}
However, this will not necessarily be the best strategy, since the optimal strategy to play against all opponents may be some compromise between the best strategy against each of them. 

Furthermore, as the agent plays against opponents, it can make guesses about their policy. The optimal policy must also balance competing factors of playing moves that refine knowledge of the current opponent, avoid driving the opponent to an unfavorable state during exploration, and optimize reward given the knowledge of the opponent.

This can be seen more clearly by optimizing over the entire policy space ($\pi^* = \argmax_{\pi \in \Pi} \E[V^\pi(s_0,\pi')]$) and breaking the policy into two components $\pi^{*}_k$ and $\pi^{*s_k}_{n-k}$, those played for the first $k$ moves, and those for the rest of the tournament, given state $s_k$ after $k$ moves.

Any policy can be broken down in this way, since information at previous states is always available at subsequent states, and the opponent's decisions has no direct knowledge of our own policy, only the mutually observable states. 

First, we approach the policy for the second set of moves given  $s_k$:
\begin{equation}
  \label{indiv_second_policy}
  \pi^{*s_k}_{\nmk} = \argmax_{\pi_{\nmk}} \E_{\pi'}[V^{\pi_{\nmk}}(s_k,\pi'_{\nmk})|s_k]  = \argmax_{\pi_{\nmk}} \sum_{\pi'_{\nmk}} V^{\pi_{\nmk}}(s_k,\pi'_{\nmk})P(\pi'_{\nmk} | s_k) 
\end{equation}
The optimal first set of moves has policy that balances receiving the optimal reward and establishing the best $s_k$ for future rewards.
\begin{align}
  \pi^{*}_k = & \argmax_{\pi_k} \E_{\pi'} \left[\E_{s_k} \left[r_{s_k} + V^{\pi^{*s_k}_\nmk}(s_k,,\pi'_\nmk)\right]\right]\\
  \label{indiv_first_policy}
  = &  \argmax_{\pi_k} \left( \sum_{\pi'} \sum_{s_k} r_{s_k}P(s_k|\pi_k,\pi'_k)P(\pi')\right. + \left. \sum_{\pi'} V^{\pi_\nmk^{*s_k}}(s_k,\pi'_\nmk)P(\pi'|s_k) \right)
\end{align}
That is, the optimal policy can be thought of as balancing exploring by tightening the posterior distribution of $P(\pi'|s_k,\pi)$, maximizing the reward $r_{s_k}$ while doing so, and maintaining the state in favorable region of the opponent's policy by maximizing $V(s_k,\pi')$. 

\subsection{Tournament Play}
\label{tournament}
We can extend the analysis from before to apply to tournament play as well. In the winner-take-all tournament, only the player with the highest  score receives any payout, and the payout received is the sum of all other players accumulated score. Formally, if the opponents are  $K = {\pi_1,\pi_2,...\pi_N}$, then $\pi^*$  maximizes the payoff as defined by the reward structure:
\begin{equation}
  \pi^* =  \argmax_\pi \sum_i f(\pi,\pi_i)\mathbb{I}[f(\pi,\pi_i) > f(\pi_i,\pi) \forall \pi_i \in K]
  \end{equation}
    \subsection{Team Play}
  \label{sec:teamplay}
This is similar to tournament play, but in this case we maximize two strategies
\begin{equation}
  \pi^*,\pi^{*'} = \argmax_{\pi,\pi'} \sum_i f(\pi,\pi_i) \mathbb{I}[f(\pi,\pi_i) > f(\pi_i,\pi) \forall \pi_i \in K]
\end{equation}

This is similar to the single agent tournament play in the previous section, but the subtle change of allowing for the simultaneous optimization of two policies  introduces much more complex dynamics into the game.

\subsection{Learning policies using Q-Learning}
To find the optimal policy, we need a way to explore the $O(2^n)$ policy space. We can do this using Q-learning, which learns a value function for the value of each action when taken at a given state: $\{\mathcal{S},\mathcal{A}\} \rightarrow \mathcal{R}$.

This function can be learned using an iterative approach to learn $\hat{Q}$, our approximation of $Q$, which corresponds with the optimal policy for the second set of moves (Eqn~\ref{indiv_second_policy}.)
\begin{equation}
  \hat{Q}(s,a) \approx Q(s,a) = \max_\pi ~ \E_{\pi'}\left[\E_a' \left[ V^{\pi}(s|| \{a,a'\},\pi')~|~\pi'\,\right]\,|\,s\,\right]
\end{equation}
We will learn this using an iterative approach, in which after each training episode, we update our Q function using the following update rule.
  \begin{equation}
    Q^{(i)}(s,a) = (1-\alpha)Q^{(i-1)}(s,a) + \alpha\left(R(s,a) + \gamma  \max_{a'} Q^{(i-1)}(s',a')\right)
  \end{equation}
  The reward function $R(s,a)$, maps state-action pairs to the the real numbers, under the constraint that the sum of rewards for a given path through the tournament would create a reward as described in the winner-take-all dynamics in \ref{tournament}. That is if $\mathcal{L}$ is a set of state-action pairs from an IPD tournament,
\begin{equation} \label{eq:reward}
  \hspace{-7mm}
  \textstyle\sum\limits_{s,a \in \mathcal{L}} R(s,a) =
  \begin{cases} \textstyle\sum\limits_{\pi_1}\sum\limits_{\pi_2 \neq \pi}V^{\pi_1}(s_0,\pi_2)  &\mbox{if} \sum\limits_{\pi'\neq \pi}  V^{\pi}(s_k,\pi') >  \sum\limits_{\tilde{\pi} \neq \pi'}  V^{\pi}(s_0,\pi') \forall \pi' \neq \pi \\
    0  & \mbox{otherwise}
  \end{cases}
\end{equation}
  Given this constraint, we will note that the reward function does not affect the optimal policy, however it could have an impact on the learning algorithm, and how well it can find the optimal policy. We will formulate this reward function as state-by-state payoffs for each bout of the iterated prisoner's dilemma, and then if the opponent loses the tournament it forfeits its accumulated payoff to the winner, and if it wins it acquires the opponents accumulated payoff in the final round.

  The policy that chosen at state $a$ is $\argmax_\alpha Q(s,a)$ with probability $1-\epsilon$ and randomly selected from $\mathcal{A}$ with probability $\epsilon$.

  \begin{wraptable}{r}{7cm}
  \begin{tabular}{{p{0.2\linewidth} p{0.7\linewidth}}}
    \toprule
    Name & Description\\
    \midrule
    Tit-for-tat & Mirror the opponent's last action\\
    \midrule
    Tit-for-two-tats & If the opponent defected in the past two bouts, defect, else cooperate\\
     \midrule
    Grudge & Cooperate until opponent defects, then defect forever\\
    \midrule
    Defector & Defect\\
    \midrule
    Cooperater & Cooperate \\
    \bottomrule
  \end{tabular}
  \caption{Strategies in tournament against which our agents learn.}  \label{tbl:opps}
\end{wraptable}

We set $\epsilon, \alpha$ proportional to the training episode raised to the $\tfrac 3 4$, where the learning rate has a minimum value of $0.03$.  To aid in convergence of the Q-table approximation, we employ experience replay, using the state-action-reward values 5 times for each training episode. For the collusion case we alternate the training of each agent, one agent updates it's parameters for 300 consecutive episodes, and then they remain fixed while the other agent updates for 300 episodes. When an agent has fixed its parameters it still employs an $\epsilon$-greedy strategy where $\epsilon$ is the lesser of the $\epsilon$ according to the learning schedule and 0.01.

\subsection{Implementation}

In order to simulate our Iterated Prisoner's Dilemma game, and compare against standard competitive agents, we used the Axelrod Python library~\cite{knight2016}. This provides many standard agents for us to use as competitors, so we are able to compare against state-of-the-art results and published tournaments. Given a set of agents, the library runs the tournament and outputs the overall scores of each agents and the moves they made, onto which we impose the winner-take-all dynamic and then compute rewards and updates as dictated by our learning methods. In our experiments we use a tournament of five opponents described in table~\ref{tbl:opps}.

\section{Results}

\begin{figure}
  \begin{minipage}{.5\linewidth}
    \includegraphics[width=1\linewidth]{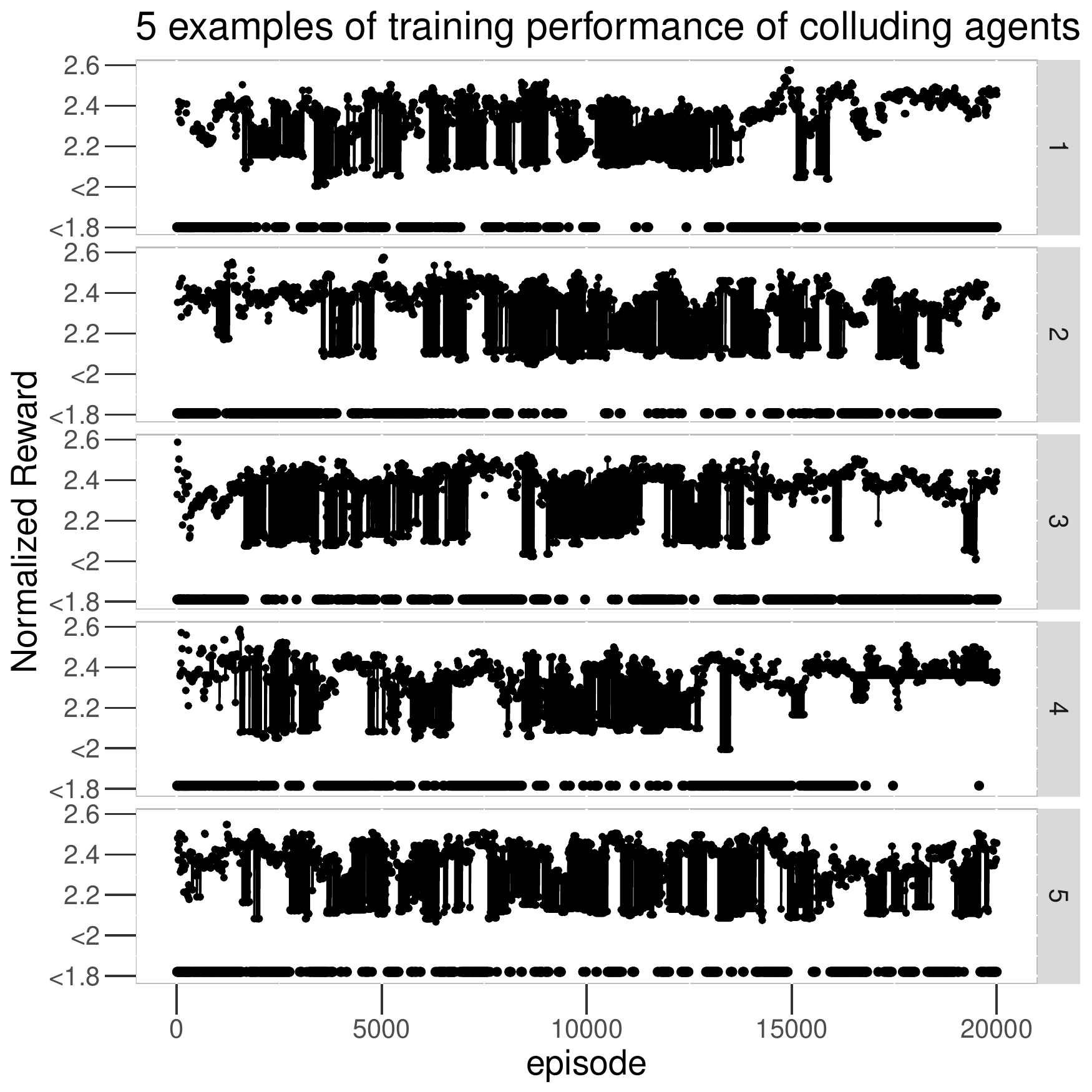}
  \caption{A sample of five teams of agents training over 20,000 episodes to play winner-take-all IPD. Due to the optimization of opponents strategies in winner-take-all dynamics, the agent often switches between winning and losing solutions over training.}
  \label{fig:strategy_learn}
\end{minipage}
\hfill\vline\hfill
\begin{minipage}{.45\linewidth}
  \includegraphics[width=0.8\linewidth]{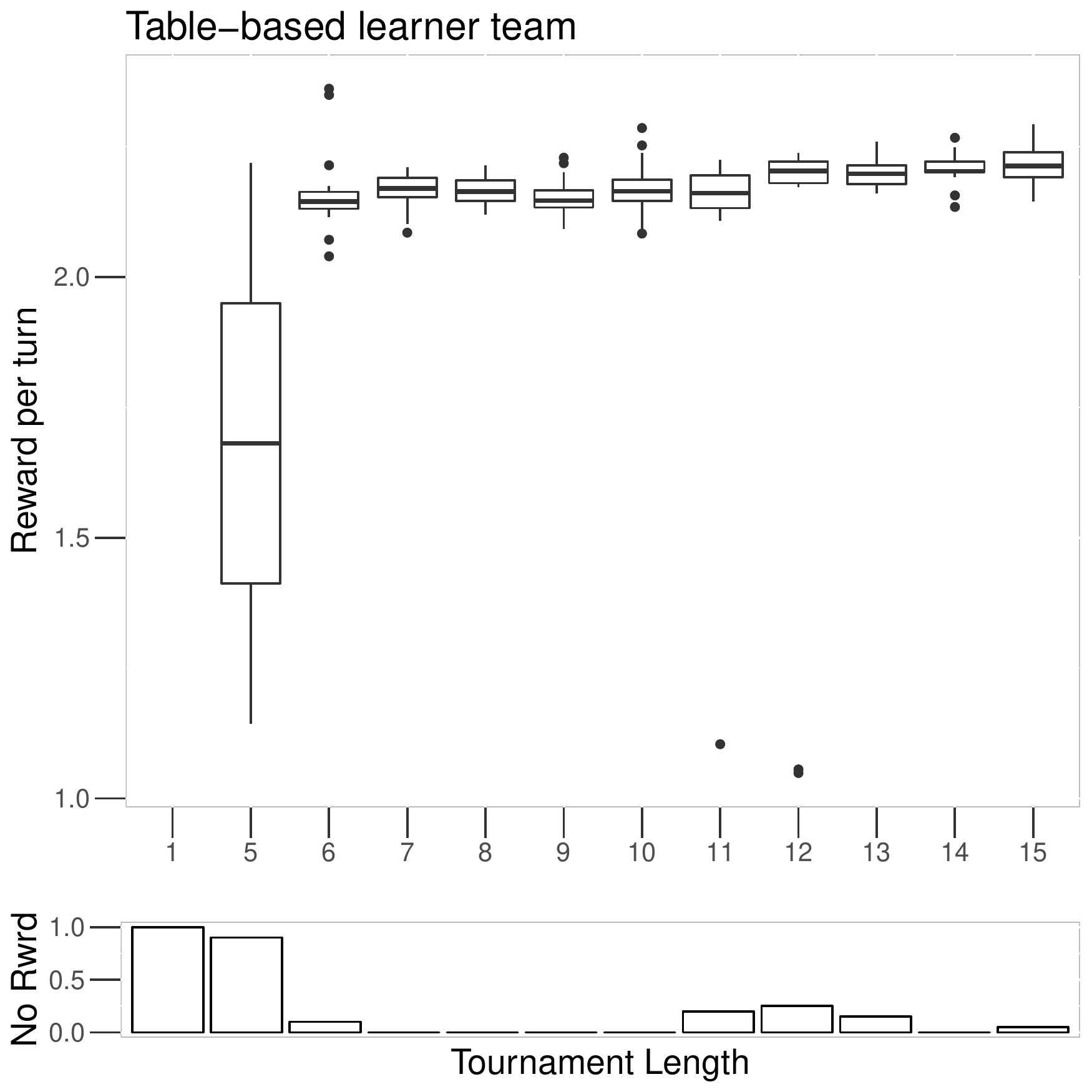}
   \caption{The median performance of the last 10\% of training episodes in the colluding tournament. For tournaments of length longer than 5, the reinforcement learning algorithm is able to reliably find winning strategies, with minimal degradation as tournament length increases.}
\label{fig:tournament_length}  
\end{minipage}
\end{figure}

In training these agents, the Q-learning algorithm performs a near-exhaustive search of the state-space, where the master and slave agent estimates Q-values for 92\% and 96\% of possible states, respectively. Furthermore, given the array of opponents we use, the agents can learn a rather trivial, non-colluding strategy of defecting on the last round. To force more interesting strategies, we impose a 3 point per opponent handicap on the the learning agents, which offsets their advantage from being able to learn the tournament length.

\begin{figure}
  \begin{center}
    \includegraphics[width=1\linewidth]{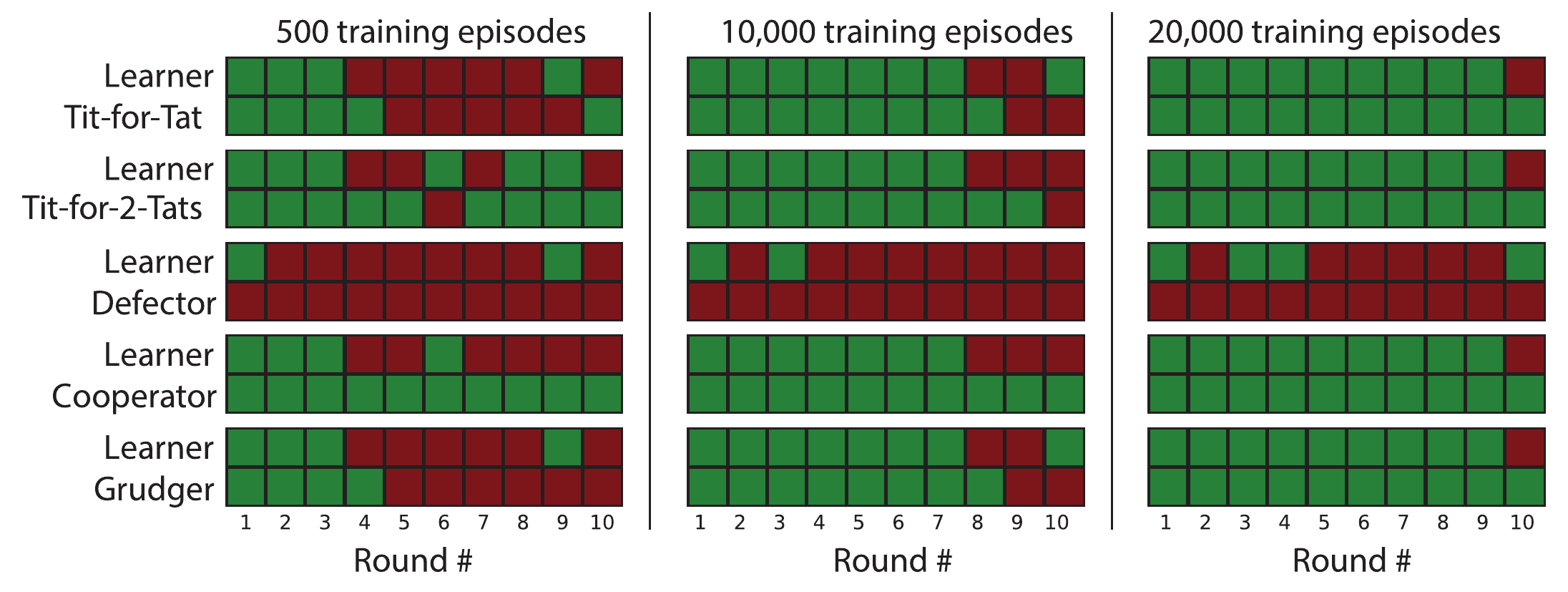}
  \end{center}

  \caption{A single agent learns an optimal strategy of 20,000 training episode. Shown here is the strategy at episodes 500, 10,000 and 20,000. Each cell in the table represents a match-up between the two opponents, where columns are bouts of the tournament and red indicates defection and green indicates cooperation. The agent first learns a hostile strategy that allows it to win, and then refines that strategy by permitting other agents to do better and maximize the total reward.}
\label{fig:strategy}
\end{figure}

The agents can learn a winning strategy for the 6-bout tournament within the first 500 training episodes, and this strategy is refined over the remainder of the training period. Without the new learner, the winner of the tournament would be tit-for-tat. As a result, the learner starts with a hostile strategy that cooperates with tit-for-tat just enough to win, but not so much that tit-for-tat could win. Over time, it learns a less aggressive strategy that maximizes the total reward by cooperating with the tit-for-tat strategy and capitulating occasionally to the defector (Fig. ~\ref{fig:strategy}).

The reward function places the optimal strategy near a critical tipping point which exacerbates the non-stationarity; once a team has the highest score in the tournament, it further improves its score by allowing other agents to perform better and minimizing the difference between its score and its competitors. Thus slight instabilities in the learned policy can cause the team to go from winning to losing (Fig.~\ref{fig:strategy_learn}). However, despite this, the agents are able to learn colluding strategies for tournaments of length greater than 5, with little degradation in performance as tournament length increases, as shown in Fig.~\ref{fig:tournament_length}.

\begin{figure}
  \includegraphics[width=0.9\linewidth]{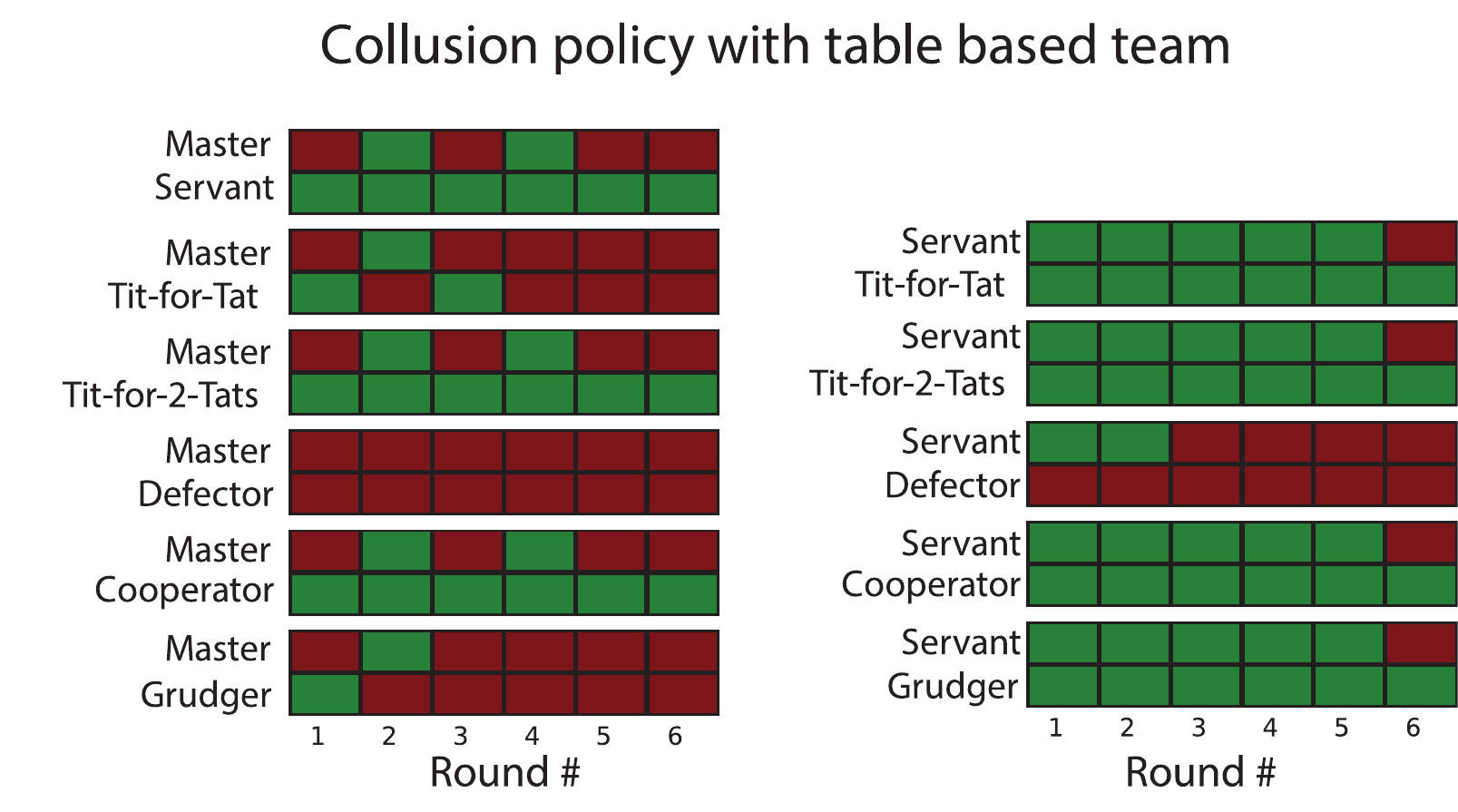}
\caption{The learned strategy for teams of colluding agents. Cooperate is in green and defect in red. Both teams learn an identification sequence and subsequently divide labor into master and servant roles, where the servant colludes despite the master's defection so that the master can win the tournament and share the cumulative reward.}
\label{fig:team_strategy}
\end{figure}

We can analyze these learned strategies to better understand a coordination protocol that the agents have learned. An example of such a winning strategy is shown in Fig. ~\ref{fig:team_strategy}. We see that the agents on the team have developed very different strategies, in which one primarily defects and becomes the winner, and the other primarily cooperates, transferring reward to its partner, enabling the partner to win, while still defecting on non-team members enough to prevent them from winning, recapitulating stylized facts of real-world implicit collusion~\cite{ivaldi2003}. Furthermore, we see that the master agent plays the same strategy against the servant as it does against the Tit-for-2-Tats, whereas the servant always cooperates when playing the master, but not with any other agent.

These interactions can be interpreted within the framework of Horn's Q and R communication principles~\cite{horn1984}. That is the Q principle, of make your contribution sufficient, and say as much as you can given R, and the R principle, make your contribution necessary and say no more than you must given Q. Thus we can see that when the master plays the first move, it completely communicates its identity in the first two moves, which is a sufficient communication, and necessary to avoid the servant defecting. However the servant, by not defecting after being defected on in the first bout, has communicated that it is either the Servant, Tit-for-2-Tats or Cooperator, which is a sufficient communication for the remainder of the strategy.

\begin{wrapfigure}{r}{6cm}
  \includegraphics[width=0.9\linewidth]{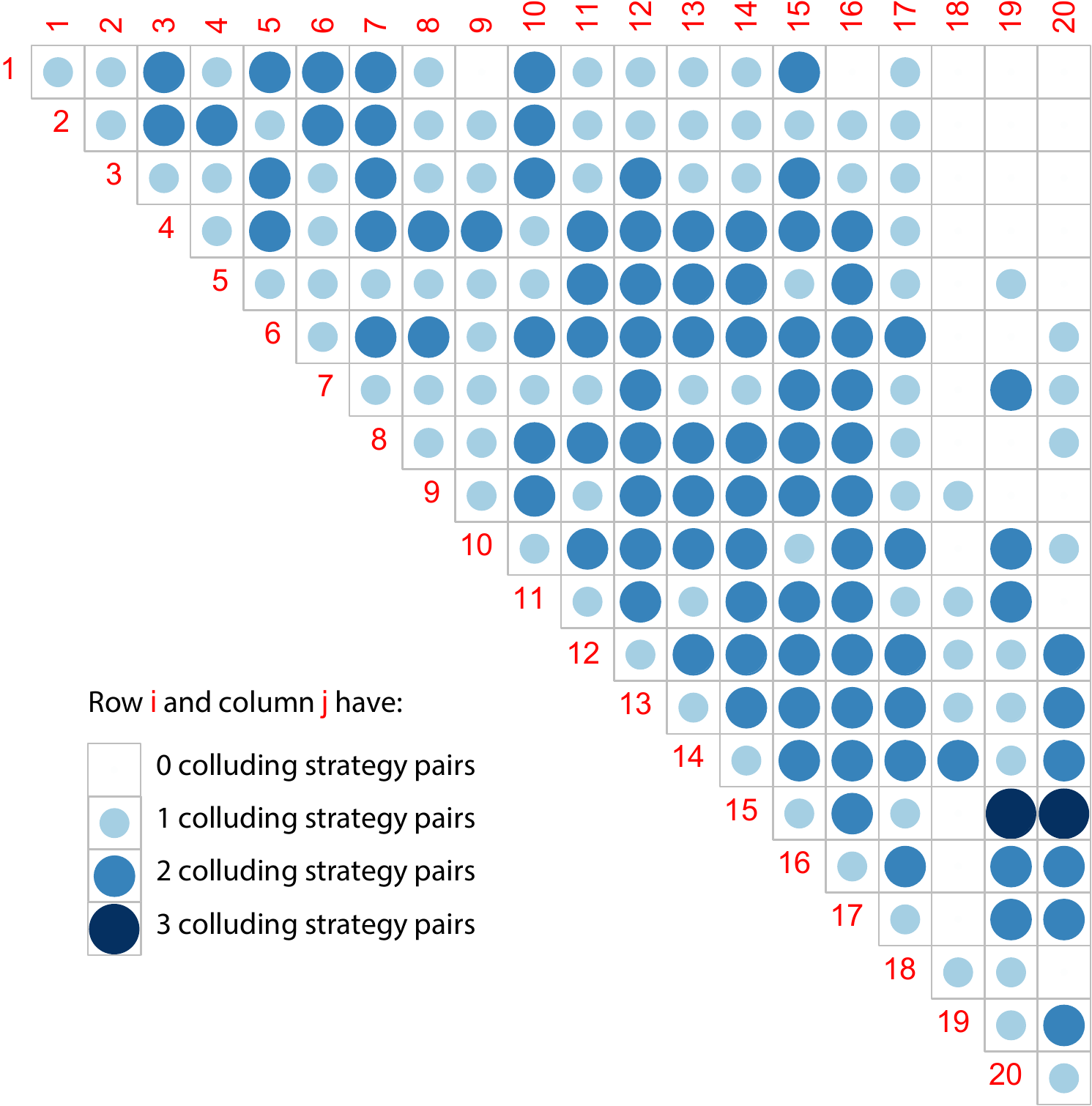}
  \caption{Agents trained independently in the same environment are able to collude with each other.\\}
\label{fig:team_cor}
\end{wrapfigure}
Despite the apparent instability within the training regime, we see that the learned coordination protocols are remarkably consistent and non-arbitrary.
 In establishing a coordination protocol, it is sometimes the case that the specifics of the signal are unimportant, but rather, that each agent has associated that particular signal with a signified behavior.   In which case, we would expect it to be unlikely for agents from different training regimes to communicate with each other. However, we find that out of 20 independent training regimes 85\% of regimes had at least one subset of master/servant pair that establish coordination strategies that win the tournament, and of those, 58\% of training regime pairs had both master/servant that could win the tournament. Surprisingly, two regimes had 3 pairs of strategies, suggesting that at least agent was able to switch between the master and servant role, depending on the context (Fig:~\ref{fig:team_cor}).

All winning learned strategies share a characteristic pattern: the first two bouts are used as a role identification sequence followed by a subsequent division of labor into a master/servant relationship. The identification sequence is Master/Servant D/C then C/C or D. This signature response is unique to the master, and the servant then cooperates regardless, while the master defects.  In this example, the master agent then proceeds to continuously defect against tit-for-tat, and play an optimal strategy against tit-for-2-tats. Although the agents could identify each other if the identification protocol was reversed, in all of the learned replicates, the identification protocol was tied with the roles, since the wrong action by either party against a non-colluder would have an adverse impact on subsequent division of labor.

\section{Conclusion}
By using the discrete state and action space of the iterated prisoner's dilemma, we can explore how strategies adapt in winner-take-all environments. In the case of a single agent, we see that an agent first learns an aggressive approach to ensure that it outperforms other agents, but over the training period cooperates more to maximize the total overall reward. 

The team of agents in this study are able to establish a strategy to communicate with each other to establish identity even when no explicit channel is present, and no information is shared before play. Simply by having a shared reward, through repeated play an identity recognition protocol and division of labor between the two agents emerge.  Furthermore, this division of labor is consistent across runs, and does not result in overfitting as has been observed in some multi-agent reinforcement learning environments~\cite{lanctot2017}.

Simply, by being trained in the same environment, agents are, more often than not, able to recognize other agents with a complementary coordination protocol. The consistency of the learned policy to adapt to the environment suggests the possibility for similar regularities may arise in implicit coordination in other domains.

\bibliography{zotero}
\bibliographystyle{unsrt}

\end{document}